\documentclass[]{article}
\makeatletter\if@twocolumn\PassOptionsToPackage{switch}{lineno}\else\fi\makeatother

\usepackage{amsfonts,amssymb,amsbsy,latexsym,amsmath,tabulary,graphicx,times,caption,fancyhdr}
\usepackage[utf8]{inputenc}

\usepackage{url,multirow,morefloats,floatflt,cancel,tfrupee}
\makeatletter

\AtBeginDocument{\@ifpackageloaded{textcomp}{}{\usepackage{textcomp}}}
\makeatother
\usepackage{colortbl}
\usepackage{xcolor}
\usepackage{pifont}
\usepackage[nointegrals]{wasysym}
\makeatletter

\def\mcWidth#1{\csname TY@F#1\endcsname+\tabcolsep}

\def\cAlignHack{\rightskip\@flushglue\leftskip\@flushglue\parindent\z@\parfillskip\z@skip}
\def\rAlignHack{\rightskip\z@skip\leftskip\@flushglue \parindent\z@\parfillskip\z@skip}

\@ifundefined{etal}{}{}

\usepackage{ifxetex}
\ifxetex\else\if@twocolumn\@ifpackageloaded{stfloats}{}{\usepackage{dblfloatfix}}\fi\fi

\AtBeginDocument{
\expandafter\ifx\csname eqalign\endcsname\relax
\def\eqalign#1{\null\vcenter{\def\\{\cr}\openup\jot\m@th
  \ialign{\strut$\displaystyle{##}$\hfil&$\displaystyle{{}##}$\hfil
      \crcr#1\crcr}}\,}
\fi
}

\AtBeginDocument{%
  \@ifpackageloaded{endfloat}%
   {\renewcommand\efloat@iwrite[1]{\immediate\expandafter\protected@write\csname efloat@post#1\endcsname{}}}{\newif\ifefloat@tables}%
}%

\def\BreakURLText#1{\@tfor\brk@tempa:=#1\do{\brk@tempa\hskip0pt}}
\let\lt=<
\let\gt=>
\def\processVert{\ifmmode|\else\textbar\fi}

\@ifundefined{subparagraph}{
\def\subparagraph{\@startsection{paragraph}{5}{2\parindent}{0ex plus 0.1ex minus 0.1ex}%
{0ex}{\normalfont\small\itshape}}%
}{}

\newcommand\role[1]{\unskip}
\newcommand\aucollab[1]{\unskip}
  
\@ifundefined{tsGraphicsScaleX}{\gdef\tsGraphicsScaleX{1}}{}
\@ifundefined{tsGraphicsScaleY}{\gdef\tsGraphicsScaleY{.9}}{}
\def\checkGraphicsWidth{\ifdim\Gin@nat@width>\linewidth
	\tsGraphicsScaleX\linewidth\else\Gin@nat@width\fi}

\def\checkGraphicsHeight{\ifdim\Gin@nat@height>.9\textheight
	\tsGraphicsScaleY\textheight\else\Gin@nat@height\fi}

\def\fixFloatSize#1{}
\let\ts@includegraphics\includegraphics

\def\inlinegraphic[#1]#2{{\edef\@tempa{#1}\edef\baseline@shift{\ifx\@tempa\@empty0\else#1\fi}\edef\tempZ{\the\numexpr(\numexpr(\baseline@shift*\f@size/100))}\protect\raisebox{\tempZ pt}{\ts@includegraphics{#2}}}}

\AtBeginDocument{\def\includegraphics{\@ifnextchar[{\ts@includegraphics}{\ts@includegraphics[width=\checkGraphicsWidth,height=\checkGraphicsHeight,keepaspectratio]}}}

\DeclareMathAlphabet{\mathpzc}{OT1}{pzc}{m}{it}

\def\URL#1#2{\@ifundefined{href}{#2}{\href{#1}{#2}}}

\def\UrlOrds{\do\*\do\-\do\~\do\'\do\"\do\-}%
\g@addto@macro{\UrlBreaks}{\UrlOrds}

\edef\fntEncoding{\f@encoding}

\makeatother

\newif\ifmultipleabstract\multipleabstractfalse%
\makeatletter

\def\wileyIndent{1pt}
\usepackage[paperheight=10in,paperwidth=6.5in,margin=2cm,headsep=.5cm,top=2.5cm,headheight=1cm]{geometry}

\renewenvironment{abstract}
{\vspace*{-1pc}\trivlist\item[]\leftskip\wileyIndent\hrulefill\par\vskip4pt\noindent\textbf{\abstractname}\mbox{\null}\\}{\par\noindent\hrulefill\endtrivlist}

\usepackage[]{footmisc}

\def\author#1{\gdef\@author{\hskip-\dimexpr(\tabcolsep)\hskip\wileyIndent\parbox{\dimexpr\textwidth-\wileyIndent}{\centering\bfseries#1}}}

\def\title#1{\linespread{1}\gdef\@title{\centering\bfseries\ifx\@articleType\@empty\else\@articleType\\\fi#1}}

\let\@articleType\@empty \def\articletype#1{\gdef\@articleType{{\normalfont\itshape#1}}}

\AtBeginDocument{\fancypagestyle{headings}{\fancyhf{}\fancyhead[C]{\RunningHead}\fancyhead[R]{\thepage}}\pagestyle{headings}}

 \def\audegree#1{}

\date{}

\makeatother

\usepackage[T1]{fontenc}
\makeatother
\usepackage[numbers,sort&compress]{natbib}

\def\thanksspace{{\phantom{\textsuperscript{\thefootnote}}}}
\begin{document}

\title{Generalizable synthetic MRI with physics-informed convolutional networks}
\author{Luuk~Jacobs\textsuperscript{1,2}, Stefano~Mandija\textsuperscript{1,2}, Hongyan~Liu\textsuperscript{1,2}, Cornelis A.T.~van den Berg\textsuperscript{1,2}, Alessandro~Sbrizzi\textsuperscript{1,2}\space and Matteo~Maspero\textsuperscript{1,2}\thanks{Corresponding author. E-mail:                     
                    m.maspero@umcutrecht.nl, matteo.maspero.it@gmail.com}\thanksspace~\\[-3pt]\normalsize\normalfont  \itshape ~\\
\textsuperscript{1}{Department of Radiotherapy\unskip, UMC Utrecht\unskip, Utrecht\unskip, The Netherlands}~\\
\textsuperscript{2}{Computational Imaging Group for MR Diagnostics and Therapy\unskip, UMC Utrecht\unskip, Utrecht\unskip, The Netherlands}}

\def\RunningHead{}\def\RunningAuthor{Jacobs \MakeLowercase{\textit{et al.}} }

\maketitle


\begin{abstract}
In this study, we develop a physics-informed deep learning-based method to synthesize multiple brain magnetic resonance imaging (MRI) contrasts from a single five-minute acquisition and investigate its ability to generalize to arbitrary contrasts to accelerate neuroimaging protocols. A dataset of fifty-five subjects acquired with a standard MRI protocol and a five-minute transient-state sequence was used to develop a physics-informed deep learning-based method. The model, based on a generative adversarial network, maps data acquired from the five-minute scan to “effective” quantitative parameter maps, here named q*-maps, by using its generated PD, T\textsubscript{1}, and T\textsubscript{2} values in a signal model to synthesize four standard contrasts (proton density-weighted, T\textsubscript{1}-weighted, T\textsubscript{2}-weighted, and T\textsubscript{2}-weighted fluid-attenuated inversion recovery), from which losses are computed. The q*-maps are compared to literature values and the synthetic contrasts are compared to an end-to-end deep learning-based method proposed by literature. The generalizability of the proposed method is investigated for five volunteers by synthesizing three non-standard contrasts unseen during training and comparing these to respective ground truth acquisitions via contrast-to-noise ratio and quantitative assessment. The physics-informed method was able to match the high-quality synthMRI of the end-to-end method for the four standard contrasts, with mean $\pm$ standard deviation structural similarity metrics above $0.75\pm0.08$ and peak signal-to-noise ratios above $22.4\pm1.9$ and $22.6\pm2.1$. Additionally, the physics-informed method provided retrospective contrast adjustment, with visually similar signal contrast and comparable contrast-to-noise ratios to the ground truth acquisitions for three sequences unused for model training, demonstrating its generalizability and potential application to accelerate neuroimaging protocols.

\def\keywordstitle{Keywords}

\end{abstract}



\section{Introduction}\label{sec1}
Conventionally, multiple complementary contrast-weighted magnetic resonance imaging (MRI) images are separately acquired for disease diagnosis. For example, standard brain imaging examinations include PD-weighted (PDw), T\textsubscript{1}-weighted (T\textsubscript{1}w), T\textsubscript{2}-weighted (T\textsubscript{2}w), and T\textsubscript{2}-weighted ﬂuid-attenuated inversion recovery (T\textsubscript{2}-FLAIR) contrasts among many other acquisitions \cite{Langen2017AdvancesImaging}. The qualitative nature of conventional MRI impedes the comparison of signal intensities between examinations from different time points, patients, and vendors. Although different scanners output qualitatively similar images, they exhibit considerable variation when viewed quantitatively \cite{Cashmore2021ClinicalMetrology}. Over the last decades, attempts to make MRI quantitative have been pursued. Quantitative MRI (qMRI) aims to facilitate standardized MRI-based measurements, reduce bias, and increase reproducibility  \cite{MargaretCheng2012PracticalRelaxometry}. In the last decades, efforts have been dedicated to developing qMRI techniques mapping biophysical tissue parameters such as T\textsubscript{1}, T\textsubscript{2}, and PD (q-maps) within a clinically viable acquisition and reconstruction time, i.e., 3-5 minutes or less. Examples of other tissue parameters are T\textsubscript{2}*, apparent diffusion coefficients, ﬂow, perfusion, and stiffness, but these are not considered in this work. Three examples of such fast qMRI techniques are MR ﬁngerprinting (MRF) \cite{Ma2013MagneticFingerprinting}, Magnetic Resonance Spin TomogrAphy in Time-domain (MR-STAT) \cite{Sbrizzi2018FastProblem}, and fitting of a signal model to multi-dynamic multi-echo (MDME) MRI \cite{Warntjes2008RapidUsage}.

To still provide radiologists with the standard contrasts routinely used for neurological diagnoses, the reconstructed q-maps can be used to generate so-called “synthetic MRI” (synthMRI), making use of signal models \cite{Ji2020SyntheticNeuroradiology}. This way, synthMRI provides intrinsically registered, multi-contrast images from a single scan, reducing examination time compared to conventional MRI. MRF \cite{Jiang2015MRReadout, Mehta2018ImageMORF}, MR-STAT \cite{Mandija2020, Kleinloog2022SyntheticTrial}, and model fitting to MDME data \cite{Warntjes2008RapidUsage} have all been explored to facilitate synthMRI. The latter commercial solution has even been implemented in clinics. Several studies showed that this commercial solution may facilitate patients’ diagnosis on par with conventional contrasts \cite{Ryu2020InitialStudy}, with applications for gliomas \cite{Blystad2017QuantitativeGliomas}, brain metastases \cite{Hagiwara2016Contrast-enhancedMetastases}, Sturge-Weber syndrome \cite{andica2016advantage}, multiple sclerosis \cite{Hagiwara2017SyntheticPlaques}, and stroke \cite{Andre2022SyntheticStudy.}. However, synthetic T\textsubscript{2}-FLAIR contrasts remain challenging for synthMRI \cite{gulani2004towards, Redpath2014MagneticSequence}. The quality of synthetic T\textsubscript{2}-FLAIR contrasts can be hindered by oversimpliﬁed signal models that do not model effects like partial volume, magnetization transfer, and ﬂow \cite{Granberg2016ClinicalStudy, Hagiwara2017SyMRIMeasurement, Hagiwara2017SyntheticPlaques}. The subsequent artifacts can result in coarse hyperintensities and a lack of contrast between the lesion and surrounding tissues, making it necessary to acquire an additional conventional T\textsubscript{2}-FLAIR to conﬁrm the diagnosis and lose the promised decrease in scan time from synthMRI \cite{Tanenbaum2017SyntheticTrial}. 

Deep learning (DL) is a subfield of machine learning that focuses on developing models that learn abstract data representations using data-driven training strategies \cite{Lecun2015DeepLearning}. DL has achieved great success for various medical imaging tasks such as segmentation \cite{Ronneberger2015U-net:Segmentation}, MRI reconstruction \cite{Wang2016AcceleratingLearning}, super-resolution \cite{Chaudhari2018Super-resolutionLearning}, and image modality conversion \cite{Wolterink2017DeepData} and has recently been proposed to facilitate synthMRI. Generative adversarial networks (GANs) \cite{goodfellow2014generative} are a class of DL that has shown promising results for a variety of medical applications, particularly for image synthesis and image-to-image translation \cite{Yi2019GenerativeReview}. Specifically, due to its fast inference times and representation capability, it has the potential to overcome imperfections in qMRI reconstructions and signal models in a data-driven manner \cite{wang2020high, Wang2020SynthesizeModel}. So far, only DL-based synthMRI methods that exclusively synthesize specific contrasts provided during training have been extensively investigated. These methods cannot generalize to unseen contrasts during training, meaning additional acquisitions are still needed if contrast tweaking or non-standard contrasts are necessary to answer clinical demands. 

This work\footnote{Code will be made publicly available on \url{https://gitlab.com/computational-imaging-lab/qstarMRI}} investigates the use of DL to obtain synthMRI from a single full-brain, five-minute scan for four routinely acquired contrasts and generalizability to unseen contrasts during training. We propose incorporating a physics-based signal model into the framework to achieve this generalizability. We will compare the physics-informed DL-based synthMRI constrasts to separately acquired ground truths and investigate whether generalizability to contrasts unseen during training for a volunteer is feasible. 

\begin{figure*}[t]
\centerline{\includegraphics[width=\textwidth]{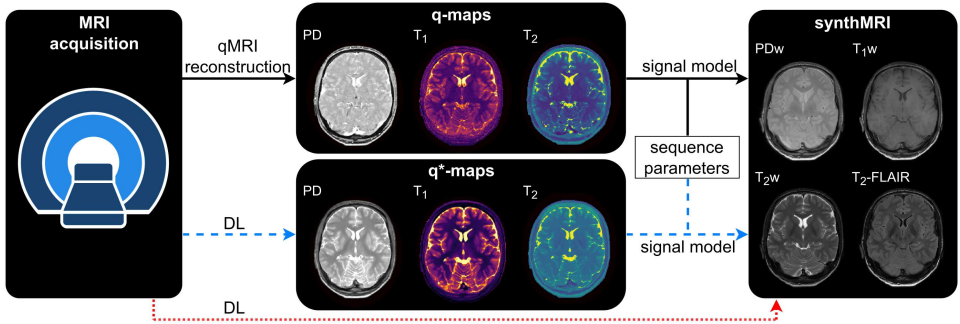}} 
\caption{\textbf{Schematic representation of possible synthMRI approaches.} The standard synthMRI approach (solid black arrows) starts with qMRI reconstruction, from which synthMRI is obtained via a signal model. End-to-end DL-based methods (dotted red arrow) have been proposed to skip the qMRI reconstruction and signal model. Our proposed physics-informed method (striped blue arrows) aims to address the lack of generalizability of the end-to-end approach by outputting effective q-maps (q*-maps) and feeding these to the signal model to obtain synthMRI.}\label{fig1}
\end{figure*}

\section{Theory}\label{sec2}
The standard synthMRI approach (Fig. \ref{fig1}, solid black arrows) consists of a qMRI reconstruction that results in q-maps and a voxel-wise signal model to synthesize contrasts from the q-maps. Various qMRI techniques have been proposed, and generally, an analytical solution to the Bloch equations is used as the signal model \cite{Jiang2015MRReadout, Mandija2020, Warntjes2008RapidUsage}, as described by: 
\begin{equation} \label{eq:TSE}
\mathrm{S}=\mathrm{PD} \cdot e^{-\mathrm{TE} / \mathrm{T}_{2}} \cdot \frac{1-\left[1-\cos \left(\theta\right)\right] e^{-\mathrm{TI} / \mathrm{T}_{1}} -\cos \left(\theta\right) e^{-\mathrm{TD} / \mathrm{T}_{1}}}{1-\cos \left(\mathrm{FA}\right) \cos \left(\theta\right) e^{-\mathrm{TD} / \mathrm{T}_{1}}}
\end{equation}
\noindent with echo time $\mathrm{TE}$, saturation pulse angle $\theta$ ($\theta=180^{\circ}$ for inversion recovery (IR) sequences, otherwise $\theta = 0^{\circ}$), inversion time $\mathrm{TI}$, delay time $\mathrm{TD}$ ($\mathrm{TD}=\mathrm{TR}-\mathrm{ETL}\cdot\mathrm{ESP}$ for turbo-spin echo (TSE) sequence, $\mathrm{TD}=\mathrm{TR}$ for spin-echo (SE) sequence, with ETL = echo train length and ESP = echo spacing), and flip angle $\mathrm{FA}$ \cite{Warntjes2008RapidUsage, Meara2005EvolutionSequences}.

Recently, supervised end-to-end DL-based methods (Fig. \ref{fig1}, dotted red arrow) have been proposed \cite{wang2020high, Wang2020SynthesizeModel}, which train a neural network (NN) to directly synthesize contrasts from the acquired data by optimizing the following objective function:
\begin{equation}
\boldsymbol{w^{*}}=\arg \min _{\boldsymbol{w}} \mathcal{L}\big(\boldsymbol{y}, \text{NN}(\boldsymbol{x}, \boldsymbol{w})\big)
\end{equation}
\noindent with weights $\boldsymbol{w}$, loss function $\mathcal{L}$, contrasts $\boldsymbol{y}$, and acquired data $\boldsymbol{x}$. For example, K. Wang et al. \cite{wang2020high} proposed synthesizing T\textsubscript{1}w, T\textsubscript{2}w, and T\textsubscript{2}-FLAIR contrasts from MRF acquisition data using DL, allowing the model to learn physiological effects in a data-driven manner. However, this restricts the synthesis during inference to only the contrasts $\boldsymbol{y}$, which are part of the paired dataset. Synthesizing different contrasts would require additional MRI scans to expand the dataset, which can be very expensive \cite{Lundervold2019AnMRI}. Furthermore, the NN would need to be retrained and re-evaluated with the new dataset.

To provide generalizability to unseen contrasts for DL-based synthMRI methods, we propose a physics-informed method (Fig. \ref{fig1}, striped blue arrows). The model first maps the acquired data to quantitative PD, T\textsubscript{1}, and T\textsubscript{2} values, so-called “effective q-maps” (q*-maps), which are then fed into the signal model described by Eq. \eqref{eq:TSE}. The model is trained to optimize for the following objective function:
\begin{equation}
\boldsymbol{w^{*}}=\arg \min _{\boldsymbol{w}} \mathcal{L}\bigg(\boldsymbol{y}, \text{S}\big(\text{NN}(\boldsymbol{x}, \boldsymbol{w}), \text{TE}, \text{TR}, \text{TI}\big)\bigg)
\end{equation}  
\noindent By performing contrast synthesis via q*-maps, all resulting synthMRI contrasts are enforced to be jointly consistent with the signal model. Similar to the standard synthMRI methods, a computationally inexpensive signal model is used for synthesis. However, our physics-informed DL-based network is trained to incorporate the signal model imperfections; effects that the signal model does not incorporate can still be captured in a data-driven manner via the q*-maps. Because of the effective nature of the q*-maps, their quantitative PD, T\textsubscript{1}, and T\textsubscript{2} values may differ from q-maps reconstructed using standard qMRI methods, which could harm synthMRI quality. We hypothesize that with this physics-informed training setup, we can use the q*-maps to provide generalizability to unseen contrasts during inference by varying the desired sequence parameters in the signal model. Additionally, contrast synthesis via q*-maps is more interpretable compared to the "black box" end-to-end methods, which improves trust and conduces clinical adaption.   

\begin{table*}[t]
\footnotesize
\caption{\textbf{Sequence parameters of the MR-STAT acquisition, standard contrasts, and unseen contrasts.} For all sequences, geometric parameters were kept constant: field of view of $224$x$224$x$134$~mm\textsuperscript{3}, acquisition and reconstruction resolution of $1$x$1$x$3$~mm\textsuperscript{3}, and $30$ slices with $1.5$~mm gaps for all scans.}
\centering
\begin{tabular}{p{0.12\linewidth}p{0.18\linewidth}p{0.04\linewidth}p{0.04\linewidth}p{0.08\linewidth}p{0.06\linewidth}p{0.04\linewidth}p{0.04\linewidth}p{0.07\linewidth}}
\hline
 & \textit{Acquisition name}  & \textit{TR} & \textit{TE} & \textit{TI} & \textit{FA} & \textit{ETL} & \textit{ESP} & \textit{TA} \\
 &  & [\textit{ms}] & [\textit{ms}] & [\textit{ms}] & [\textit{degrees}] & [-] & [\textit{ms}] & [\textit{min:sec}] \\
\hline

\textbf{MR-STAT acquisition} & Multi2D spoiled Cartesian GRE & 8.9 & 4.7 & N/A & 0-90 & N/A & N/A & 4:50   \\
\hline
\textbf{Standard} & T\textsubscript{1}w MS SE & 864   & 14  & N/A  & 70 & 1  & N/A & 3:16 \\
\textbf{contrasts}& T\textsubscript{2}w MS TSE      & 3254  & 80  & N/A  & 90 & 15  & 10 & 1:44 \\
& PDw MS TSE                      & 2800  & 20  & N/A  & 90 & 14 & 9 & 1:41 \\
& T\textsubscript{2}-FLAIR MS TSE & 10000 & 120 & 2800 & 90 & 24 & 9.6 & 3:40 \\
\hline

\textbf{Unseen} & T\textsubscript{1}/T\textsubscript{2}w MS TSE & 2000 & 50 & N/A & 70 & 14 & 6.7 & 3:00\\
\textbf{contrasts}& TI400 MS TSE & 10000 & 120  & 400  & 90 & 24 & 9.6 & 3:40\\
& DIR MS TSE & 10608 & 25 & 325/3400* & 90 & 17 & 6.8 & 5:39 \\

\hline
\noalign{\vskip 0.5mm}   
\end{tabular}

\textbf{Abbrevriations: }GRE = gradient-echo; MS = multi-slice; DIR = double inversion recovery; TSE = turbo spin-echo; TR = repetition time; TE = echo time; TI = inversion time; FA = flip angle; ETL = echo train length; ESP = echo spacing; TA = acquisition time. *Two inversion pulses are applied with different TIs 

\label{table:sequence param}
\end{table*}

\section{Methods}\label{sec3}
\subsection{Data acquisition and processing}
In this study, we collected data from forty patients (brain tumor (11), stroke (10), epilepsy (10), and multiple sclerosis (9)) and ten volunteers (10) included in an institutional review board-approved study to assess the clinical usability of MR-STAT-based synthMRI \cite{Kleinloog2022SyntheticTrial}. Subjects were imaged on a 3~T scanner (Ingenia MR-RT, Philips, Best, The Netherlands) and 15-channel head coil (HeadSpine, Philips, Best, The Netherlands). The acquired sequences included MR-STAT \cite{Sbrizzi2018FastProblem, vanderHeide2020High-resolutionAlgorithm} and four standard neuro contrast acquisitions (PDw, T\textsubscript{1}w, T\textsubscript{2}w, and T\textsubscript{2}-FLAIR), whose imaging protocols are summarized in Tab. \ref{table:sequence param}. The MR-STAT sequence is a five-minute transient-state multi-2D spoiled gradient-echo with a slowly varying flip angle preceded by a non-selective 180$^{\circ}$ inversion pulse. Flip angle variations were subdivided into five acquisition subsets, obtaining five separate k-spaces. Two stroke patients were excluded due to severe motion artifacts in the standard contrast acquisitions. To assess the feasibility of the physics-informed approach to synthesize unseen contrasts, five additional volunteers were imaged with the MR-STAT, four standard contrasts, and three different sequences with increasing deviation from the standard contrasts to assess the framework flexibility (also summarized in Tab. \ref{table:sequence param}): (i) T\textsubscript{12}w as an intermediate contrast between the T\textsubscript{1}w and T\textsubscript{2}w contrasts; (ii) TI400 as a T\textsubscript{2}w IR sequence like the T\textsubscript{2}-FLAIR contrast, but with a different TI; (iii) a double IR (DIR) \cite{turetschek1998double} as a completely new sequence.

The 15-channel MR-STAT data (each capturing ﬁve k-spaces) were compressed into a single virtual coil using singular value decomposition \cite{golub2013matrix}. For pre-processing purposes, q-maps were reconstructed using the MR-STAT framework \cite{Sbrizzi2018FastProblem}, from which “MR-STAT contrasts” (PDw, T\textsubscript{1}w, T\textsubscript{2}w, T\textsubscript{2}-FLAIR) were synthesized using the signal model \cite{Mandija2020}. White matter (WM), gray matter (GM), and cerebrospinal ﬂuid (CSF) were segmented from the MR-STAT contrasts using the predeﬁned tissue probability maps in SPM12 (v7771) \cite{penny2011statistical}. All other pre-processing steps were performed using Python (v3.7.4) \cite{10.5555/1593511}. The conventional contrasts were normalized by scaling the WM intensity mode to the average mode among all subjects and intensities were clipped based on estimated values that clipped most of the highest skull intensities. To simplify the contrast synthesis starting from the image domain, the ﬁve complex k-spaces of the virtual coil were fast Fourier transformed, creating five complex “fast Fourier transform (FFT) images”. The magnitudes of the FFT images were min-max scaled to [0, 1] per volume, and the real and imaginary components were used as separate real-valued input channels. We compared using raw FFT as input against using MR-STAT q-maps in a preliminary analysis reported in the supplementary materials, demonstrating that starting from raw FFT is beneﬁcial for synthesis. Based on this study, only investigations using raw FFT images as input are therefore reported. Finally, the conventional contrasts were rigidly registered per slice to their respective MR-STAT contrast that is intrinsically registered to the FFT images since these are generated from the same k-spaces. The registration was performed with SimpleElastix (v2.0.0rc2) \cite{Marstal_2016_CVPR_Workshops}, adopting an adaptive stochastic gradient descent optimizer \cite{Klein2009AdaptiveRegistration}, the Mattes mutual information similarity metric \cite{Mattes2001NonrigidRegistration}, and a recursive pyramid strategy with isotropic smoothing and down-sampling of 8, 4, 2, and 1 for the four resolutions.

\begin{figure*}[t]
\centerline{\includegraphics[width=\textwidth]{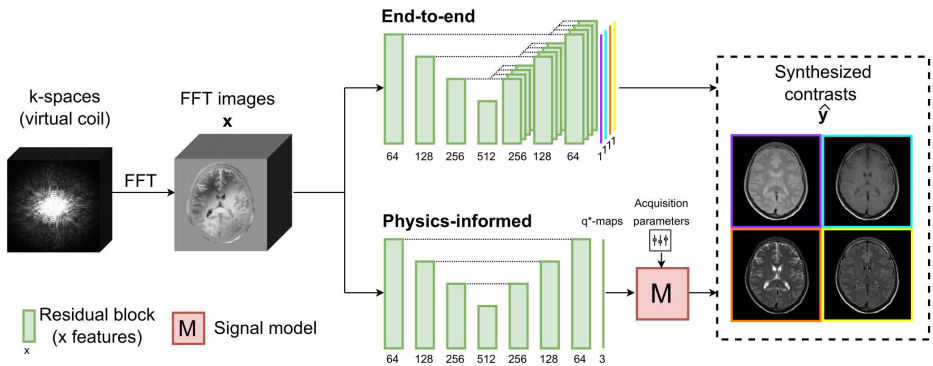}} 
\caption{\textbf{Schematic representation of the synthMRI methods.} The ﬁve complex FFT images ($\textbf{x}$) were fed into the two architectures to output four contrasts ($\hat{\textbf{y}}$) directly (end-to-end) or via q*-maps followed by the physics model M (physics-informed). }\label{fig2}
\end{figure*}

\subsection{Deep learning synthesis} \label{sec:dls}
\noindent Two models were adopted: 1) end-to-end and 2) physics-informed (proposed). GANs were employed for both methods due to their success in various medical image translation tasks \cite{Yi2019GenerativeReview}. Specifically, conditional GANs (cGANs) \cite{mirza2014conditional} were used, where either a synthesized or conventional contrast was concatenated to the FFT images for conditioning \cite{pix2pix}. The architectures of the generator and discriminators of the end-to-end and physics-informed approaches were kept similar (Fig. \ref{fig2}), consisting of a ResUNet generator \cite{Zhang2018RoadU-Net} with three down-sampling operations (33M and 13M parameters, respectively) and four contrast-specific conditional PatchGAN discriminators \cite{pix2pix} (2.8M parameters each). The end-to-end approach mapped the FFT images to the four conventional contrasts using a separate decoder (one output channel) for each contrast, similar to the work by K. Wang et al.  \cite{wang2020high}. Contrarily, the physics-informed method mapped the FFT images to the q*-maps using a single decode (three output channels) from which contrasts were synthesized in a voxel-wise manner using the signal model described in Eq. \eqref{eq:TSE}. The resulting signal value is min-max scaled to [0, 1] per slice and used as the signal intensities. Synthetic T\textsubscript{1} and T\textsubscript{2} q*-maps are clipped to 5 and 1.5~s, respectively, after observing that these values were never exceeded in the training population. All networks used the Kaiming weight initialization \cite{He2015DelvingClassification} and rectified linear unit (ReLU) activations \cite{Nair2010RectifiedMachines}, where the leaky variant \cite{maas2013rectifier} with slope 0.2 was adopted for the discriminators. As described in \eqref{cont_loss}, the content loss compares the synthesized contrasts $\hat{y}$ to the conventional ones $y$, for each contrast $i$. 
\begin{equation} \label{cont_loss}
\mathcal{L}_{\textit{cont, i}} = \lambda_{1}\|\hat{y}_i-y_i\|_{1}+\lambda_{2}\|\hat{y}_i-y_i\|_{2}  -\lambda_{3} \cdot \operatorname{SSIM}(\hat{y}_i, y_i) +\lambda_{4}\left\|\left(\phi(\hat{y}_i)-\phi(y_i)\right)\right\|_{2}
\end{equation}

\noindent The voxel-wise terms comprised a weighted linear combination of L1 and L2 losses. A structural similarity metric (SSIM) \cite{Wang2004ImageSimilarity} loss was added to incorporate local structures and optimize perceptual quality. To further optimize the perceptual similarity between the generated and conventional contrasts, we also minimized the L2-norm of the features extracted from the 14\textsuperscript{th} convolutional layer of a pre-trained VGG-19 net ($\phi$) \cite{Simonyan2015VeryRecognition}. Adding the least-squares adversarial losses gives us the full objective functions for the generator $\mathrm{G}$ and discriminators $\mathrm{D}_{i}$, described by 
\begin{equation} \label{g_loss}
\min _{G} \sum_{i=1}^{4} \Big(\gamma_{1} \mathbb{E}\big[(1 - D_i(x, \hat{y}_i))^{2}\big] + \mathcal{L}_{\textit{cont, i}}\Big)
\end{equation}
\noindent and
\begin{equation} \label{di_loss}
\min _{D_i} \gamma_{2} \mathbb{E}\big[D_i(x, \hat{y}_i)^{2}\big]+\gamma_{3} \mathbb{E}\big[(1 - D_i(x, y_i))^{2}\big]
\end{equation}
\noindent respectively \cite{Mao2017LeastNetworks}.

Five-fold cross-validation was performed, randomly splitting the 48 subjects (30 slices each) \cite{Kleinloog2022SyntheticTrial} into 34, 5, and 9 subjects for training, validation, and testing, respectively. All folds have an equal distribution of pathologies in the training and test sets. Hyperparameter optimization was performed on a single fold containing only healthy volunteers in the validation set. The training was repeated for the four remaining folds (with 35 and 8 subjects in the train and test set, respectively, for two of the folds) using these same hyperparameters. Each subject is in the test set once, except for the ﬁve held-out validation subjects, which were only used for hyperparameter optimization, giving a total of 43 subjects in the test set.

The loss weights, types of augmentations, and batch size were manually optimized for the physics-informed approach and were adopted for the end-to-end approach. For this optimization, the SSIM and a sharpness estimation (the variance of the image convolved with a 3x3 Laplacian kernel) \cite{Pech-Pacheco2000DiatomStudy} were used as metrics, where SSIM was the primary metric considered when SSIM and sharpness disagreed. The discriminators were updated twice per generator update and weights [$\lambda_1, \lambda_2, \lambda_3, \lambda_4$] and [$\gamma_1, \gamma_2, \gamma_3$] were set to [1, 1, 4, 0.005] and [0.05, 0.025, 0.025]. During training, the images were augmented by applying flipping and translation ($\pm20$ voxels), both horizontally and vertically, and rotation ($\pm15^{\circ}$). In preliminary experiments, we observed that unhealthy tissues were underrepresented compared to healthy tissue during training. To counteract this imbalance, we doubled the chance that a slice with pathology could occur. All methods were implemented in PyTorch v1.9.0 \cite{Paszke2019PyTorch:Library} and training with a batch size of 1 took about 38 and 26 hours for the end-to-end and physics-informed approach, respectively, on an Nvidia Tesla P100 GPU and Intel E5-2690v4 CPUs.

A cosine learning rate decay over 300 epochs, with a restart after 100 epochs to prevent overfitting, was adopted for the generator and discriminators, where the ﬁnal epoch was always used for evaluation. The initial learning rate values were determined using separate automatic Bayesian searches for the physics-informed and end-to-end approaches. These optimizations, performed in Weight\&Biases \cite{wandb}, minimized the sum of the content losses as deﬁned in Eq. \eqref{cont_loss} for learning rates between 5e-5 and 1e-2 (logarithmic uniform distribution) using ten searches. The optimal learning rate for the physics-informed and end-to-end approaches were 1e-3 and 2.5e-3, respectively.

\subsection{Experiments}
\subsubsection{Synthetic q*-maps}
The values of test volunteer’s q*-maps were compared to literature values for WM and GM \cite{Bojorquez2017WhatT, Hagiwara2019LinearityControls}. Due to the signal equation’s linear dependency on PD (see Eq. \eqref{eq:TSE}) and the applied min-max scaling for contrast synthesis (see Section \ref{sec:dls}), the PD q*-maps were not considered quantitative and were compared to literature using the GM/WM ratio instead, to eliminate the effect of different scaling factors.

\subsubsection{Synthetic contrasts}
To assess whether the physics-informed method is able to match the performance of the end-to-end approach for the conventional contrasts, quantitative and qualitative comparisons were performed. Firstly, the physics-informed method was quantitatively compared to the end-to-end method in terms of the test set’s SSIM and peak signal-to-noise ratio (PSNR) values. The lower and upper five slices were left out for this and all upcoming experiments due to increased noise in the conventional contrasts. A Shapiro-Wilk test \cite{Shapiro1965AnSamples} showed that the image similarity metrics were not normally distributed. Therefore, a non-parametric paired Wilcoxon signed-rank test \cite{Wilcoxon1945IndividualMethods} was used to compare the physics-informed method to the end-to-end method (two-tailed with a signiﬁcance level of $p \leq 0.05$), where the mean SSIM and PSNR of each subject were considered to satisfy the independent sample assumption. All statistical analyses were performed using SciPy (v1.3.1) \cite{virtanen2020scipy} and statsmodels (v0.13.2) \cite{seabold2010statsmodels}. Secondly, a visual comparison is shown for a multiple sclerosis patient to assess robustness of the methods synthesizing lesions of various shapes and sizes.

A more detailed qualitative comparison was performed focusing on T\textsubscript{2}-FLAIR contrasts, as these have been reported to be the most challenging contrasts \cite{Tanenbaum2017SyntheticTrial}: The occurrence of motion artifacts, CSF flow, checkerboard artifacts, and unrealistic lesion suppression in T\textsubscript{2}-FLAIR contrasts were reported, as first classified by a non-clinical intern and revised by an experienced MRI physicist, together with example images of the observed artifacts. Finally, the mean $\pm$ standard deviation inference times over all the test sets to obtain synthMRI for a single subject were reported for both methods.

To assess the generalizability of the physics-informed approach to unseen contrasts, contrasts from three additional sequences were synthesized (Tab. \ref{table:sequence param}) for five volunteers (see Tab. \ref{table:sequence param}): (i) T\textsubscript{12}w, (ii) TI400, and (iii) DIR. The DIR contrast was synthesized using the following signal model:
\begin{equation}
\mathrm{S}=\mathrm{PD} \cdot e^{-\mathrm{TE} / \mathrm{T}_{2}} \cdot\left(1-2 e^{-\mathrm{TI}_{1} / \mathrm{T}_{1}}+2 e^{-\mathrm{TI}_{2} /\mathrm{T}_{1}}-e^{-\mathrm{TD} /\mathrm{T}_{1}}\right)
\end{equation}

\noindent with $\mathrm{TI}_{1}$ the time between the first $180^{\circ}$ pulse and the excitation pulse (suppressing CSF) and $\mathrm{TI}_{2}$ the time between the second $180^{\circ}$ pulse and the excitation pulse (suppressing WM). The synthetic contrasts were skull-stripped and visually compared to separately acquired ground truth acquisitions. For quantitative image comparison, the mean $\pm$ standard deviation contrast-to-noise ratios (CNRs) were compared for five test volunteers, defined as the mean signal intensity difference between two regions of interest divided by the image noise (estimated by computing the standard deviation of intensities in a manually selected homogeneous square patch of WM). The regions of interest were chosen to be GM-WM for T\textsubscript{12}w, CSF-WM for TI400, and GM-WM for DIR.  


\begin{figure*}[t]
\centerline{\includegraphics[width=\textwidth]{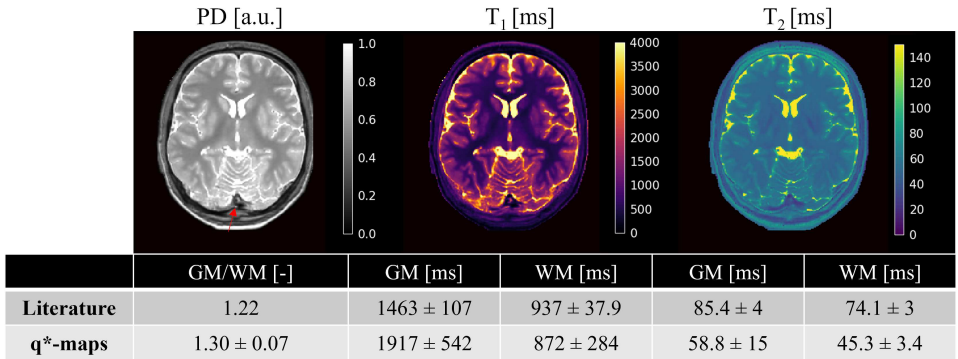}} 
\caption{\textbf{Q*-maps for a volunteer.} A transverse slice example of the q*-maps for a volunteer. The GM/WM ratio is compared to Ref. \cite{Hagiwara2019LinearityControls} and the T\textsubscript{1} and T\textsubscript{2} values to the median value in Ref. \cite{Bojorquez2017WhatT}. The standard deviations of the PD ratio and T\textsubscript{1} and T\textsubscript{2} values describe inter-subject and intra-subject variability, respectively. The PD q*-maps and MR-STAT maps are normalized to the 75\textsuperscript{th} percentile of the CSF volume for visualization purposes. The superior sagittal sinus is highlighted in the PD maps.}\label{fig3}
\end{figure*}

\section{Results}\label{sec4}
\subsection{Synthetic q*-maps}
The PD GM/WM ratios of the q*-maps are close to the ratio reported in literature \cite{Hagiwara2019LinearityControls} and present lower values where there are inﬂow effects, such as in the superior sagittal sinus (Fig. \ref{fig3}). The T\textsubscript{1} values in WM are close to the reported values, whereas the values in GM differ substantially \cite{Bojorquez2017WhatT}. The T\textsubscript{2} values of WM and GM are underestimated compared to the literature values \cite{Bojorquez2017WhatT}.


\begin{figure*}[t]
\centerline{\includegraphics[width=\textwidth]{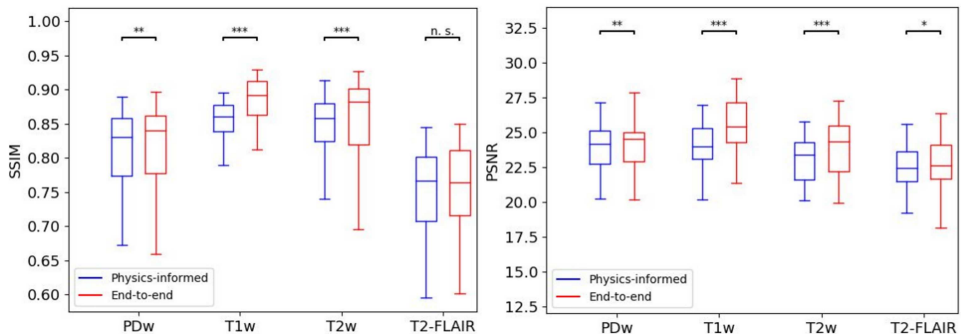}} 
\caption{\textbf{Quantitative comparison of the end-to-end and physics-informed approach.} The SSIM and PSNR of the contrasts synthesized using the end-to-end (blue) and physics-informed (red) approach are visualized using boxplots and statistically compared. *($0.01 < p \leq 0.05$) **($0.001 < p \leq 0.01$) ***($p \leq 0.001$) n.s. = not significant ($p > 0.05$).}\label{fig4}
\end{figure*}

\begin{figure*}[b!]
\centerline{\includegraphics[width=\textwidth]{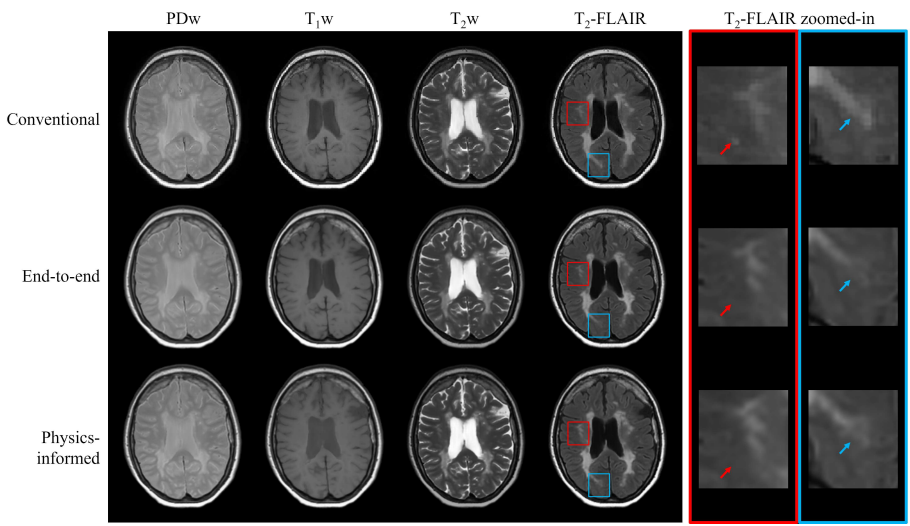}} 
\caption{\textbf{SynthMRI on multiple sclerosis.} Contrasts are shown for conventional acquisition (top row), end-to-end (middle row), and physics-informed (bottom row) approaches, where the right columns zooming in on patches of the T\textsuperscript{2}-FLAIR contrast with red and blue arrows highlighting missed and hypointense/blurred lesions, respectively.}\label{fig5}
\end{figure*}

\subsection{Synthetic contrasts}
A quantitative comparison demonstrated that the physics-informed method performed comparable to the end-to-end method for the PDw, T\textsubscript{1}w, T\textsubscript{2}w, and T\textsubscript{2}-FLAIR contrasts in terms of SSIM and PSNR (Fig. \ref{fig4}). The biggest difference in performance is in the T\textsubscript{1}w contrasts, having SSIM values of $0.88 \pm 0.03$ and $0.85 \pm 0.03$ and PSNR values of $25.5 \pm 2.0$ and $24.0 \pm 1.6$ for the end-to-end and physics-informed approaches, respectively. The inference times for all four contrasts averaged over all test subjects were $2.38 \pm 0.03$~s and $1.34 \pm 0.02$~s for the end-to-end and physics-informed approaches, respectively.

The end-to-end and physics-informed approaches do not display large visual differences, for example, as demonstrated for a multiple sclerosis patient in Fig. \ref{fig5}. The conventional and synthetic PDw and T\textsubscript{1}w images lack contrast between the lesions and the surrounding healthy tissue. The CSF of the physics-informed T\textsubscript{1}w contrast was consistently hyperintense compared to the conventional contrast. Both synthetic T\textsubscript{2}w contrasts accurately capture almost all lesions present in the conventional contrast, but for the T\textsubscript{2}-FLAIR contrasts, some smaller lesions were missed (Fig. \ref{fig5}, red arrows) and hypointense or blurred (Fig. \ref{fig5}, blue arrows).

\begin{figure*}[t]
\centerline{\includegraphics[width=\textwidth]{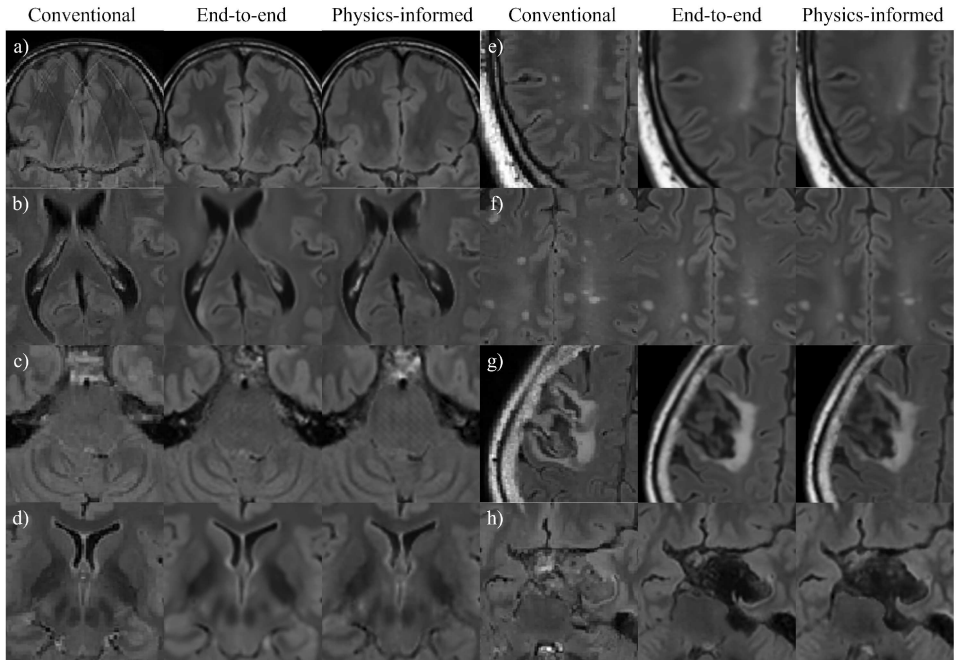}} 
\caption{\textbf{Qualitative comparison of the conventional contrasts and synthMRI methods focusing on T\textsubscript{2}-FLAIR only.} In the left three columns, zoomed-in structures of volunteers with a) motion artifacts in the conventional contrast, b) ﬂow artifacts in all contrasts, c) checkerboard artifacts in the brainstem of the synthMRI contrasts, and d) blurry basal ganglia structures in the synthMRI contrasts. In the right three columns, zoomed-in lesions are shown of patients with e) epilepsy, f) multiple sclerosis, g) stroke, and h) tumor.}\label{fig6}
\end{figure*}

Here, a more detailed analysis of the T\textsubscript{2}-FLAIR contrasts is provided, where the artifact counts are determined using manual inspection of the images. Motion artifacts were detected for 17 out of 43 test subjects (40\%) in the conventional contrasts, which were not observed for the synthMRI of any subject (Fig. \ref{fig6}a). CSF flow artifacts are present for 43 subjects (100\%) in the synthMRI (Fig. \ref{fig6}b). Grid-like (chessboard) artifacts were introduced in the brainstem for 23 (53\%) and 4 (9\%) subjects for the physics-informed and end-to-end approach, respectively (Fig. \ref{fig6}c). Similar chessboard patterns were observed in the parenchyma of 4 (9\%) subjects for the physics-informed method. Finally, synthMRI (especially the end-to-end approach) can appear blurrier than the conventional contrasts, for example, the basal ganglia structures (Fig. \ref{fig6}d). This finding has been quantitatively conﬁrmed in a study reported in the supplementary materials investigating image sharpness. Regarding pathologies, both synthMRI methods sometimes miss or result in hypointense lesions with a smaller volume, for example, in epilepsy or multiple sclerosis patients (Fig. \ref{fig6}e-f). Also, more prominent, complicated pathologies such as a stroke can result in inaccuracies. Although the gross shape is captured, hypointensities and blurriness were observed (Fig. \ref{fig6}g). Finally, both synthMRI methods were found to result in an unrealistic suppression of tumor contrast for 4 out of 10 patients (40\%) (Fig. \ref{fig6}h).

\begin{figure*}[t]
\centerline{\includegraphics[width=\textwidth]{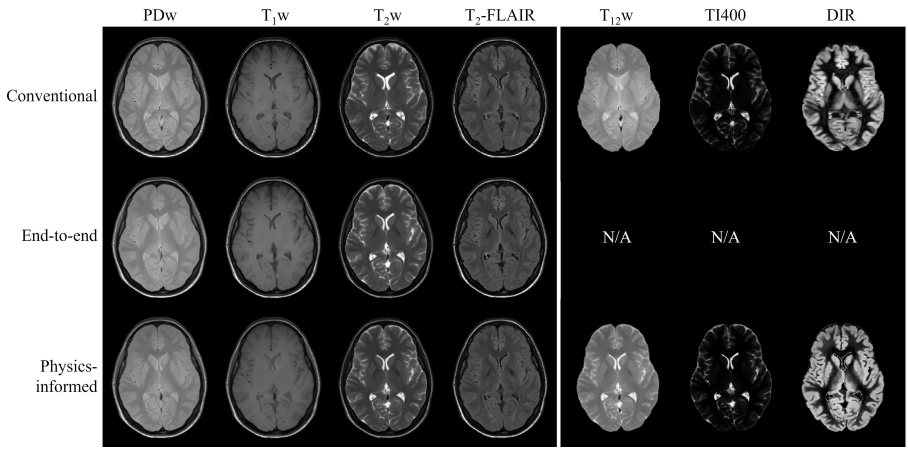}} 
\caption{\textbf{SynthMRI on a volunteer with unseen contrasts.} Standard contrasts seen during training (left) and unseen, skull-stripped contrasts (right) are shown for the conventional contrasts and the end-to-end and physics-informed approaches. Window leveling is kept identical among the same contrasts.}\label{fig7}
\end{figure*}

The standard end-to-end approach can only synthesize contrasts seen during training. Contrarily, the experiment results suggest that the physics-informed approach can synthesize any desired contrast from the q*-maps (Fig. \ref{fig7}). Similar to the T\textsubscript{1}w contrast, the physics-informed synthetic T\textsubscript{12}w has a hyperintense CSF compared to the conventional T\textsubscript{12}w contrast. For the physics-informed synthetic TI400 and DIR contrasts, the signal suppression is almost identical to the ground truth contrasts, although slight hyperintensities can be observed in the WM of the synthetic DIR contrasts. The CNR values of the synthetic T\textsubscript{12}w and TI400 contrasts were close to the values of their respective ground truths: 2.65 $\pm$ 0.12 versus 2.24 $\pm$ 0.21 and 56.9 $\pm$ 11.4 versus 54.8 $\pm$ 7.96, respectively. The CNR values of the synthetic DIR contrast were considerably lower than for its ground truth: 12.8 $\pm$ 3.31 versus 17.6 $\pm$ 1.79. 

\section{Discussion}\label{sec5}
This work investigated a physics-informed GAN-based framework synthesizing MRI from a single five-minute acquisition, obtaining four standard neurological sequences in about 2 seconds.  For the PDw, T\textsubscript{1}w, T\textsubscript{2}w, and T\textsubscript{2} FLAIR contrasts, we showed that the physics-informed method closely resembles those of the end-to-end approach, visually and in terms of SSIM and PSNR. For the T\textsubscript{2}-FLAIR contrasts, both methods could result in hypointensities for smaller lesions and more checkerboard artifacts appeared for the physics-informed approach. Future work may focus on minimizing and finding the causes of these artifacts. However, we demonstrated that the physics-informed method resulted in sharper T\textsubscript{2}-FLAIR contrasts and facilitated synthetization of unseen contrasts, with only minor visual deviations and comparable CNR values compared to the conventional, separately acquired ground truth contrasts. The inferior CNR of the synthetic DIR contrasts compared to its ground truth can be explained by the suboptimal signal suppression leading to a higher WM tissue inhomogeneity and thus a higher noise, lowering the CNR. Further work is required to evaluate the clinical relevance of these trade-offs offered by the proposed physics-informed approach.

This work is a continuation of our previous work, where the physics-informed framework was originally proposed \cite{Jacobs2022, Jacobs2022Thesis}. In this work, the framework was further developed and the evaluation was greatly expanded via cross-validation, correlation of q*-maps with baseline q-maps, quantitative artifact analysis, and a more detailed analysis of the generalizability. DL-based end-to-end methods have been proposed to improve synthMRI quality by using MRF or MDME acquisition data as input to a GAN and bypassing the oversimplified signal model \cite{wang2020high, Wang2020SynthesizeModel}. K. Wang et al. \cite{wang2020high} and G. Wang et al. \cite{Wang2020SynthesizeModel} adopted a single generator to synthesize all contrasts to exploit complementary information from diﬀerent contrasts, forming our end-to-end method. In terms of network design, the single-branch and multi-branch U-net \cite{Ronneberger2015U-net:Segmentation} architectures from K. Wang et al. \cite{wang2020high} were adapted for the physics-informed and end-to-end approaches, respectively, where we implemented residual units to simplify training \cite{Zhang2018RoadU-Net}. The real and imaginary components of the FFT images were used as separate real-valued input channels. Alternatively, complex-valued networks could be explored to accurately represent the magnitude and phase to improve the network’s performance \cite{Cole2021AnalysisApplications}. The translations performed by the generators were kept within the image domain, similar to K. Wang et al. \cite{wang2020high} and G. Wang et al. \cite{Wang2020SynthesizeModel}, preventing the need for fully-connected layers in the generator to estimate the Fourier transform, which is memory intensive and require more data to avoid overfitting \cite{Eo2020AcceleratingDirection}. Similar to our proposed framework, Pal et al. \cite{pal2022personalized} proposed to use a deep image prior to denoise three SE contrasts and fit a signal model to extract q-maps from the denoised contrasts, facilitating personalized synthMRI by omitting the need for an external training dataset. However, the high inference time is undesirable in a clinical setting and the methods were trained and evaluated using exclusively synthetic data. Denck et al. \cite{Denck2021MR-contrast-awareNetworks} proposed the ﬁrst GAN to allow retrospective contrast adjustments to synthesize missing or corrupted knee MRI contrasts from existing contrasts, where MR physics was incorporated implicitly via style transfer. Our proposed method obtains this feature by explicitly modeling MR physics, similar to the physical priors used by Moya-Sáez et al. \cite{Moya-Saez2021AData, Moya-Saez2021APriors}. Ref. \cite{Moya-Saez2021AData} was trained using exclusively synthetic data and Ref. \cite{Moya-Saez2021APriors} was finetuned using a relatively small and homogeneous in-vivo dataset, whereas we used a large cohort of in-vivo data containing heterogeneous pathologies. Furthermore, we pushed the generalizability of our method to contrasts which have not been observed by the network. This is the first GAN-based synthMRI framework allowing retrospective contrast adjustments from a single acquisition.

The previously proposed MDME-based and MRF-based synthMRI frameworks suffer from sub-optimal T\textsubscript{2}-FLAIR contrasts compared to conventional acquisitions, generally attributed to an oversimplified signal model \cite{Hagiwara2017SyntheticPlaques, Granberg2016ClinicalStudy, Hagiwara2017SyMRIMeasurement}. Signal model extensions incorporating partial volume and flow effects have shown promising results to improve the T\textsubscript{2}-FLAIR contrasts for MRF-based synthMRI \cite{cencini2019chasing, deshmane2016accurate}. Partial volume and magnetization transfer modeling could be explored for our physics-informed method. We showed that our data-driven approach already resolved flow effects without requiring explicit modeling, for example, in the superior sagittal sinus. Grid-like structures in the parenchyma were reported by G. Wang et al. \cite{Wang2020SynthesizeModel} when adding adversarial and perceptual losses. Future work should investigate the inﬂuence of these loss terms on our observed checkerboard artifacts. Although GANs may introduce hallucinations \cite{Cohen2018DistributionTranslation}, these were not observed in our methods. 

Considering that the adopted similarity metrics have been suggested not to be ideal surrogate measures of MRI quality as determined by radiologist evaluation \cite{Mason2020ComparisonImages}, a large-scale clinical study like the one performed by Kleinloog et al. \cite{Kleinloog2022SyntheticTrial} should be initiated to validate the proposed method further. Although the dataset was relatively small, a heterogeneous pathological cohort has been considered, exposing the networks to a diverse distribution of lesions, making the dataset representative of the real-world data that the network would infer in the clinic and improving generalization performance \cite{Lundervold2019AnMRI}.

Although the MR-STAT acquisition is used as input for our synthMRI framework, substituting a different acquisition that encodes sufficient information to characterize the tissue parameters (such as MRF or MDME data) would be conceptually straightforward, making the proposed framework independent of MR-STAT and generally applicable to any acquisition. Furthermore, the data were acquired on a single scanner, so inter-scanner variability should be investigated before clinical use. Currently, the q*-maps are used to facilitate synthetization of unseen contrasts. In future work, we plan to investigate the q*-maps’ diagnostic application as a fast, surrogate qMRI method to standardize MRI-based measurements, reduce bias, and increase reproducibility \cite{MargaretCheng2012PracticalRelaxometry}.

Acquiring a single, five-minute acquisition and synthesizing the contrasts reduces scan time compared to the conventional, separate contrast acquisitions. This way, synthMRI reduces examination costs, increases patient throughput, and makes MRI a more accessible imaging modality. The proposed incorporation of the signal model provides added interpretability, which is important for clinical adaptation of deep learning-based methods \cite{Lundervold2019AnMRI}. Furthermore, the provided retrospective contrast adjustment gives imaging specialists the flexibility to select the desired post-scan contrasts, which may decrease the number of patient recalls \cite{Goncalves2018SyntheticDirections}, possibly resulting in improved diagnostics. For example, it has been shown that synthesizing DIR contrasts may improve the detection of multiple sclerosis plaques \cite{Hagiwara2017SyntheticPlaques}. Additionally, a shorter scan time is advantageous for pediatric brain imaging \cite{West2017ClinicalExperience}, potentially decreasing the use of general anesthesia and preventing potential long-term complications in the developing brain of children \cite{Goncalves2018SyntheticDirections, Ing2017LatentChildhood}.


\section{Conclusion}\label{sec6}
We demonstrated that the proposed physics-informed method synthesizes high-quality standard contrasts from a single full-brain five-minute acquisition. Also, we proved the feasibility of generating synthMRI of unseen sequences during training. The proposed method is able to match the quality of PDw, T\textsubscript{1}w, T\textsubscript{2}w, and T\textsubscript{2} FLAIR contrasts of an end-to-end framework but provides additional flexibility in synthesizing additional contrasts. 


\section*{Acknowledgments}
We want to thank Sarah Jacobs for the clinical feedback. This work would not have been possible without the involvement of Anja van der Kolk, Beyza Koktas, and Sarah Jacobs, who contributed to patient selection and inclusion. We also gratefully acknowledge the support of NVIDIA Corporation with the donation of the Quadro RTX 5000 GPU used for the prototyping part of this research.

\subsection*{Author contributions}
L.J. sketched the idea, implemented the methods, and performed the analyses. M.M. participated in developing the idea. S.M., H.L., and M.M. revised the methods and analyses. C.B. and A.S. supervised and facilitated the work. All authors discussed the results and contributed to the writing of the manuscript. 

\subsection*{Financial disclosure}

None reported.

\subsection*{Conflict of interest}

The authors declare no potential conflict of interests.

\bibliography{MRM-AMA}
\bibliographystyle{wileynum}%
\vfill\pagebreak






\end{document}